\documentclass[aps,prl,reprint,superscriptaddress,floatfix
]{revtex4-1}
\usepackage{graphicx}
\usepackage{color}
\usepackage{dcolumn}
\usepackage{bm}
\usepackage{amssymb}
\usepackage{color,amsmath}
\usepackage{soul,xcolor} 
\setstcolor{red}
\usepackage{siunitx}
\usepackage{epstopdf}
\usepackage{gensymb}

\DeclareSIUnit\angstrom{\text {Å}}

\begin{document}
\title{Digital Volumetric Biopsy Cores Improve
Gleason Grading of Prostate Cancer
Using Deep Learning}

\author{Ekaterina Redekop}
\affiliation{Department of Bioengineering, University of California, Los Angeles, USA}
\author{Mara Pleasure}
\affiliation{Medical Informatics, University of California, Los Angeles, USA}
\author{Zichen Wang}
\affiliation{Department of Bioengineering, University of California, Los Angeles, USA}
\author{Anthony Sisk}
\affiliation{Department of Pathology, University of California, Los Angeles, USA}
\author{Yang Zong}
\affiliation{Department of Pathology, University of California, Los Angeles, USA}
\author{Kimberly Flores}
\affiliation{Department of Pathology, University of California, Los Angeles, USA}
\author{William Speier}

\affiliation{Department of Bioengineering, University of California, Los Angeles, USA}
\affiliation{Medical Informatics, University of California, Los Angeles, USA}
\affiliation{Department of Radiological Sciences, University of California, Los Angeles, USA}
\author{Corey W. Arnold}
\email{CWArnold@mednet.ucla.edu}
\affiliation{Department of Bioengineering, University of California, Los Angeles, USA}
\affiliation{Medical Informatics, University of California, Los Angeles, USA}
\affiliation{Department of Pathology, University of California, Los Angeles, USA}
\affiliation{Department of Radiological Sciences, University of California, Los Angeles, USA}

\begin{abstract}
\end{abstract}

\maketitle

\textbf{
Prostate cancer (PCa) was the
most frequently diagnosed cancer among American men
in 2023~\citep{siegel2023cancer}. The histological grading of biopsies is essential for diagnosis, and various deep learning-based solutions have been developed to assist with this task. Existing deep learning frameworks are typically applied to individual 2D cross-sections sliced from 3D biopsy tissue specimens. This process impedes the analysis of complex tissue structures such as glands, which can vary depending on the tissue slice examined. We propose a novel digital pathology data source called a “volumetric core,” obtained via the extraction and co-alignment of serially sectioned tissue sections using a novel morphology-preserving alignment framework. We trained an attention-based multiple-instance learning (ABMIL) framework on deep features extracted from volumetric patches to automatically classify the Gleason Grade Group (GGG). To handle volumetric patches, we used a modified video transformer with a deep feature extractor pretrained using self-supervised learning. We ran our morphology-preserving alignment framework to construct 10,210 volumetric cores, leaving out 30\% for pretraining. The rest of the dataset was used to train ABMIL, which resulted in a 0.958 macro-average AUC, 0.671 F1 score, 0.661 precision, and 0.695 recall averaged across all five GGG significantly outperforming the 2D baselines.}

\begin{figure*}[ht!]
\includegraphics[width=6.5in]{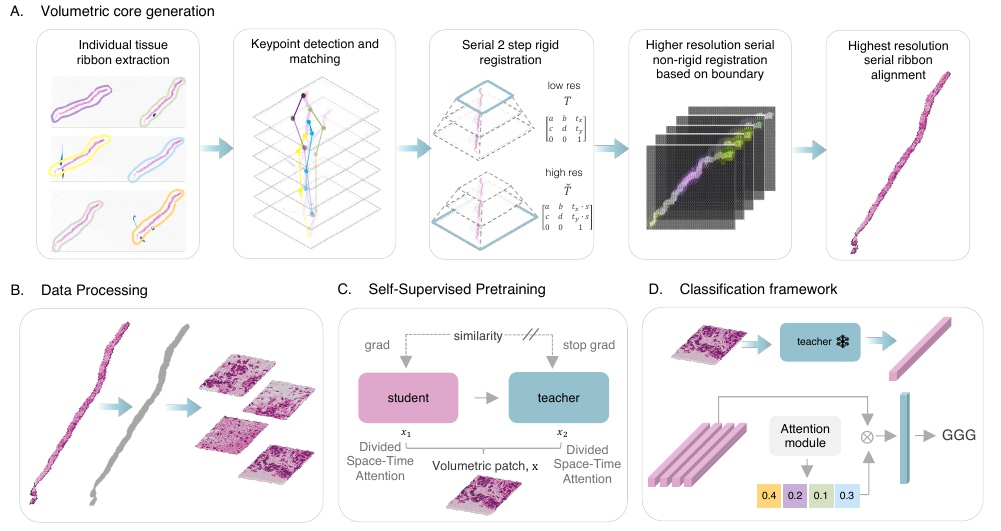}
\caption{\textbf{VCore computational workflow.}
(\textbf{A.} Morphology-preserving tissue alignment framework consists of 3 main steps: individual tissue ribbon extraction, serial rigid registration, and high-resolution non-rigid registration based on the boundary.) 
(\textbf{B.} VCore first separates the core tissue from the background and then splits it into the volumetric patches without overlap, removing patches containing less than 60\% of tissue.) 
(\textbf{C.} Self-supervised pretraining framework based on DINO with TimeSformer backbone for spatiotemporal
feature learning directly from a volumetric patch.) 
(\textbf{D.} The volumetric patches are processed with a pretrained feature encoder network. The resulting set of features is combined using a learnable attention module to produce a final patient-level prediction. )
}
\label{fig1}
\end{figure*}

\section*{\textbf{Introduction}} 
The transrectal ultrasound-guided (TRUS) prostate biopsy procedure is considered the standard for diagnosing prostate cancer~\citep{cornford2024eau}. Tissue cores are extracted systematically throughout the prostate using ultrasound guidance with or without additional magnetic resonance imaging (MRI) guidance. Following tissue processing, pathologists then use microscope-based examination to determine the presence and grade of cancer in the extracted biopsy cores. Cancer morphology and aggressiveness are rated using the Gleason scoring system (GS). The International Society of Urological Pathology (ISUP) Gleason Grade Group (GGG) system has been adopted to further categorize the GS system based on risk stratification into five categories (1-5), with increased risk of cancer mortality corresponding to increasing GGG number~\citep{epstein20162014,gleason1974prediction}. 

Biopsy tissue cores are inherently three-dimensional (3D), but routine examination requires cutting multiple serial sections from tissue blocks. This procedure potentially leads to the loss of crucial details regarding volumetric tissue morphology, cellular architecture, and the spatial distribution of pathological structures. The number of cuts can vary between laboratories, but up to 48 sections can be obtained from a single needle core~\citep{bostwick1997evaluating}. Recently, non-destructive 3D imaging technologies such as open-top light-sheet
microscopy (OTLS) have emerged to better characterize the morphology in a tissue volume~\citep{roberts2012toward, kiemen2023tissue,tanaka2017whole, glaser2017light}. However, clinical translation of these technologies is limited due to the complexity of manual evaluation, the absence of computational platforms to analyze complex 3D tissue arrays, and the maximum magnification of 10x, which is two times smaller than the resolution typically required for detailed cellular analysis. These limitations have led to the development of deep learning (DL)-based tools to process 3D pathology images and predict patient outcomes. 

MAMBA and its extension TriPath are tools designed for 3D pathology analysis using attention-based multiple instance learning (ABMIL)~\citep{ilse2018attention}. They were trained in a weakly supervised manner to predict patient-level risk for prostate cancer across two different imaging modalities: 50 simulated core needle biopsies from prostatectomy specimens imaged with OTLS and 45 prostatectomy specimens imaged with microcomputed tomography (microCT). The study has significant limitations. First, the relatively small size of the datasets—only 50 and 45 samples for OTLS and microCT, respectively—limits the generalizability and robustness of the model's predictions. Second, MAMBA and TriPath lack a pathology-specific feature encoder, crucial for capturing the intricate details needed for high-quality data representation in digital pathology. 
\begin{table}[]
\centering
  \caption{Number of cores for each GGG.}
  \label{tab:GG_number}
\begin{tabular}{lcccc}
Grade Group                  & Train & Validation & Test & Total \\ \hline
BN     & 3563  & 502        & 978  & 5043  \\
GGG 1   & 711   & 97         & 220  & 1028  \\
GGG 2   & 564   & 90         & 186  & 840   \\
GGG 3   & 272   & 49         & 89   & 410   \\
GGG 4/5 & 274   & 34         & 86   & 394   \\ \hline
\end{tabular}
\end{table}

In this work, we introduce \textbf{VCore}, Gleason Grading pipeline utilizing a novel data source - a volumetric core constructed from traditional 2D pathology scans (see FIG. \ref{fig1}). A novel morphology-preserving alignment framework is proposed to construct a volumetric core from routinely scanned 2D prostate biopsy whole slide images (WSIs) (FIG. \ref{fig1}A). Pathology image alignment has two major difficulties: 1) aligning serial tissue sections while maintaining morphological integrity, and 2) applying a complex registration process to gigapixel images. The VALIS framework~\cite {gatenbee2023virtual} was developed to address these challenges by performing automatic preprocessing, normalization, tissue detection, feature extraction, and matching for automatic tissue alignment. The SuperGlue Graph Neural Network keypoint matching framework \citep{sarlin2020superglue} was developed to improve traditional feature matching algorithms used in VALIS (RANSAC, Tukey’s approach, and neighbor match filtering) by utilizing graph neural networks to solve a differentiable optimal transport problem. Previous work has focused on performing registration with features extracted directly from the ribbon rather than solely from the ribbon boundaries, which leads to the increased risk of altering the tissues morphology~\cite {kiemen2022coda, gatenbee2023virtual,du2005efficient}. Our approach builds upon VALIS and SuperGlue and introduces a novel variation to the elastic deformation technique. Specifically, our method leverages the ribbon's boundary to perform non-rigid registration to preserve nuclei and glandular morphology. The developed co-registration framework operates on WSIs with magnification up to 20x (0.5 $\mu$m/pixel), which is higher than data obtained with novel 3D technologies~\citep{song2024analysis} and therefore provides more tissue morphology details which are essential for cancer grade prediction~\citep{li2021multi}.

As part of the VCore framework, we propose a pipeline for constructing volumetric biopsy cores, which are used to improve DL-based GGG diagnosis and microscope-based examination workflows. Our model is based on an ABMIL~\citep{ilse2018attention} with a novel volumetric feature encoder (FIG. \ref{fig1}D). We propose a modification of the widely used DINO contrastive learning framework~\citep{ba2016layer}, optimizing the feature encoder architecture to work with volumetric biopsy data (FIG. \ref{fig1}C).
We also worked with two experienced pathologists to conduct a reader study showing traditional microscope-based examination is improved by enhancing the functionality of a digital pathology slide viewer. The slide viewer allows for rapid scrolling through the volumetric core, which we compare to conventional microscope-based signout of glass slides.

\section*{\textbf{Materials and Methods}} 
\subsection*{\textbf{Data}} 
We construct our volumetric core dataset by applying an alignment framework, detailed in the next section, to 10,210 TRUS prostate biopsy cores. Each core was serially sectioned to ~6-16 ribbons and placed on three glass slides. Glass slides were stained with Hematoyxlin\&Eosin (H\&E) and scanned at 40× magnification (0.25 $\mu$m/pixel). Digitized WSIs were downsampled to 20× magnification, corresponding to 0.5$\mu$m/pixel. Tissue masks were extracted for each 2D WSI scan by converting downsampled slides into the HSV color space and thresholding the hue channel. Morphological closing was used to refine tissue masks and smooth small gaps.
After alignment, non-overlapping volumetric patches of size 256x256 with more than 60\% tissue were extracted from the slide (see FIG. \ref{fig1}B).

We randomly divided the dataset by patient into 70\% for training, 10\% for validation, and 20\% for testing. We stratified the dataset by patient-level GGG determined by the max GGG in each patient’s set of biopsy cores, which resulted in 5,384 cores for training, 772 for validation, and 1,559 for the test cohort. (Table~\ref{tab:GG_number}). 

A subset of 25 cases from the test subset was selected for clinical validation. This subset included five cases from each GGG group. 

\subsection*{\textbf{Morphology Preserving Alignment}} 
The alignment framework is built upon the VALIS framework~\citep{gatenbee2023virtual} and the SuperGlue Graph Neural Network keypoint matching framework \citep{sarlin2020superglue}. The framework consists of: 1) extracting individual tissue ribbons from each downsampled 2D WSI using morphological labeling, 2) aligning individual tissue ribbons to obtain initial components for rigid transformation, and 3) performing rigid and non-rigid registration of the corresponding high-resolution WSIs. Initial keypoints and descriptors are obtained using the scale-invariant feature transform (SIFT) key-point extractor \citep{lowe2004distinctive}. Using the set of keypoints and corresponding visual descriptors for each unregistered pair of images, the assignments are estimated by solving a differentiable optimal transport problem with costs predicted by a graph neural network \citep{sarlin2020superglue}. The estimated transformation matrix has only rotation, translation, and scaling components to eliminate tissue deterioration by non-rigid deformations. Due to the linear relationship across the WSI levels, an estimation of the translation vector at the original resolution can be calculated by multiplying the result with the
corresponding down-sampling factors. The scaling factors and the rotation angle are independent of the resolution and can be directly applied to the higher resolution levels.
Rigid registration is performed sequentially, aligning each image in the stack to the reference image. The 1st image is the reference, then the second image is aligned to it, then the third is aligned to the registered version of the second one, and so forth. Only features present in both neighboring images are used to align each image to the next one in the stack. The rigidly aligned images are stacked together to create a non-rigid registration mask. 

Subsequently, the bounding box of this mask is utilized to extract higher-resolution versions of the tissue from each slide at 20x magnification, which are employed for non-rigid registration.
To perform non-rigid registration, we find 2D displacement fields by optimizing metrics based on ribbon boundaries using the SimpleElastix method \citep{marstal2016simpleelastix}. Ribbon boundaries are obtained by finding ribbon masks from each rigidly-aligned ribbon. After the displacement fields are found, the resultant transformation is applied to the original non-binary version of the image.

\subsection*{\textbf{Slide-Level Classification}} 
Due to the gigapixel size of WSIs, patching is commonly used to train deep-learning models. However, patch-based annotations are time-consuming and expensive to obtain, instead slide-level labels which are more readily available can be used as a weak local label. In this instance, Multiple Instance Learning (MIL) can be used to represent each WSI as a 'bag' or set of instances.~\citep{campanella2019clinical}. Attention-based MIL classifies the entire bag of instances instead of individual instances by using a trainable attention module to learn the relative importance of each instance for the final prediction~\citep{ilse2018attention}. Localization is achieved by aggregating the learned attention values into a final attention map. The attention module operates on deep features extracted by a pretrained encoder. 

\subsection*{\textbf{Volumetric Pretraining}} 
\label{seq:vol_pretr}
To pretrain the volumetric feature extractor, we used the DINO framework, in which a student network was trained to match the probability distribution of a siamese teacher network using contrastive loss~\citep{ba2016layer}. While "global" (standard resolution crops) views of the image are passed through the teacher network, "local" (low-resolution crops) are passed through the student network, encouraging "local-to-global" correspondences learned through the contrastive objective. In the DINO framework, teacher and student networks share the same architecture, and the teacher is optimized using a momentum encoder.
The framework was adapted for the volumetric core by substituting a regular visual transformer (ViT)~\citep{dosovitskiy2020image}-based architecture for student and teacher networks with TimeSformer for spatiotemporal feature learning directly from a volumetric patch~\citep{bertasius2021space}. The TimeSFormer framework was built on top of standard transformer architecture to enable spatiotemporal feature learning by sequentially applying temporal and spatial attention. In this work, the temporal component is adapted to function as a depth component within the volumetric patch. Following the ViT setting, each 2D section from the volumetric patch is decomposed into $N$ non-overlapping patches of size $P\times P$, which are flattened into vectors $x_{(p,t)} \in \mathbf{R}^{3P^{2}}$, where $p = 1, ..., N$ denoting spatial location and $t=1, ..., F$ - index over z-stack of 2D slices.  Each patch $x_{(p,t)}$ is mapped into an embedding vector using learnable matric $E\in \mathbf{R}^{D\times 3P^{2}}$ and combined with a positional embedding $e_{p,t}^{pos} \in \mathbf{R}^{D}$ to encode spatiotemporal position:
\begin{equation}
z_{p,t}^{0} = Ex_{p,t} + e_{p,t}^{pos}
\end{equation}
The resulting vector represents the input to the Transformer, which consists of $L$ encoding blocks. For each block
$l$, the representation 
$z_{p,t}^{l-1}$
  from the preceding block is used to compute a query ($q$), key ($k$), and value ($v$) vector for each patch:
\begin{equation}
q_{(p,t)}^{(l,a)} = W_{Q}^{(l,a)}LN(z_{p,t}^{(l-1)}) \in  \mathbf{R}^{D_{h}}
\end{equation}
\begin{equation}
k_{(p,t)}^{(l,a)} = W_{K}^{(l,a)}LN(z_{p,t}^{(l-1)}) \in  \mathbf{R}^{D_{h}}
\end{equation}
\begin{equation}
v_{(p,t)}^{(l,a)} = W_{V}^{(l,a)}LN(z_{p,t}^{(l-1)}) \in  \mathbf{R}^{D_{h}}
\end{equation}
where $LN$ represents layer normalization~\citep{ba2016layer}, $a\in\{1,...,A\}$ is the index over attention head in a transformer block.

For divided attention, temporal attention is computed first within each block $l$:
\begin{equation}
    \alpha_{(l,a)(p,t)}^{time} = \text{SM} \left( \frac{q^{(l,a)^{T}}_{{(p,t)}}}{\sqrt{D_h}} \cdot \left\{ k^{(l,a)}_{(0,0)}, k^{(l,a)}_{(p,t')}\right\}_{t'=1,...,F} \right)
\end{equation}
where $SM$ - softmax activation function.
The encoding $z^{(l)time}_{p,t}$ is obtained by computing the weighted sum of value vectors using self-attention coefficients from each attention head. New key/query/value vectors are calculated from $z^{(l)time}_{p,t}$ and it is then passed for spatial attention computation:
\begin{equation}
    \alpha_{(l,a)(p,t)}^{space} = \text{SM} \left( \frac{q^{(l,a)^{T}}_{{(p,t)}}}{\sqrt{D_h}} \cdot \left\{ k^{(l,a)}_{(0,0)}, k^{(l,a)}_{(p',t)}\right\}_{p'=1,...,N} \right)
\end{equation}
The resulting vector $z^{(l)space}_{p,t}$ is passed to the MLP to compute the final encoding $z_{p,t}^{l}$ of the patch for the block $l$.

\subsection*{\textbf{Model Comparison}}
The proposed DINO pretraining, utilizing the volumetric core, was compared to a baseline pretraining approach that used 2D image patches derived from slicing the volumetric core along the z-axis. The baseline approach employed the default ViT backbone for pretraining. After pretraining, a frozen backbone was used to extract patch-level feature vectors, which were aggregated into a core-level vector through attention-based MIL. 

Additionally, following the MAMBA framework~\citep{song2023weakly}, we used a ResNet50-based feature extractor pretrained on video clips (Kinetics-400) followed by attention-based MIL for comparison.

\subsection*{\textbf{Clinical Validation}} 
We conducted a reader study for GGG grading based on volumetric core recruiting two experienced Genitourinary pathologists (P1 and P2). We randomly sampled a cohort of 25 patients from our test subset with equal distribution among the five GGG. In total, we organized two rounds of reader studies. 
In the initial round, both pathologists performed a conventional microscope-based examination of serial tissue sections, completing the spreadsheet as they do in their routine practice. The spreadsheet included the final diagnosis, GGG, estimated percentage of the tumor's core occupied, estimated tumor size in $mm$, cribriform, intraductal carcinoma, and perineural invasion. The pathologist selects the final diagnosis from one of the following categories: Benign, PCA (Prostate Cancer), HGPIN (High-grade prostatic intraepithelial neoplasia), ASAP (Atypical small acinar proliferation), PINATYP (High-grade PIN with adjacent small acinar proliferation), ASAP-HI (Highly suspicious for carcinoma),
AIP (Atypical intraductal proliferation, suspicious for intraductal carcinoma), BFM (Benign fibromuscular tissue, no prostatic tissue seen). Pathologists performed a similar assessment in the second round after a 3-month washout period using volumetric core samples and a digital slide viewer with the ability to scroll through consecutive and co-registered tissue sections of volumetric core developed based on OpenSeadragon~\citep{open2019web}, a JavaScript library that allows building a viewer with advanced zooming support. The improved functionality allows scrolling through an unlimited number of co-registered tiles at up to 20x magnification by shift-scroll combination. 

\subsection*{\textbf{Evaluation metrics}}
To evaluate the morphology preserving image alignment framework, registration error was calculated as the median distance ($\mu m$) between subsequent keypoints in the stack for each image. Core-based registration error was then calculated as the average of the registration errors collected for each image in the stack, weighted by the number of matched features per pair of images.

We used the AUC, multi-class weighted Precision, Recall, and F1 score metrics to evaluate the GGG prediction performance of the VCore. We utilized the McNemar statistical test to evaluate the significance of the difference in classification accuracy between the models. 

We assessed the intra- and inter-pathologist agreement within each study type (microscope-based and digital) by computing the quadratic weighted kappa metric.

We used an Attention Rollout method~\citep{abnar2020quantifying} to visualize attentions for the TimeSformer backbone of the DINO contrastive learning framework. Assuming the attention weights determine the proportion of the incoming information that can propagate through the layers, these weights can be used to approximate how the information flows between network layers. If $A_l$ is a 2D attention weight matrix at layer $l$, $A_l[i,j]$ would represent the attention of token $i$ at layer $l$ to token $j$ from layer $l-1$. The attention to the input tokens from the volumetric patch is computed by recursively multiplying the attention weights matrices, starting from the input layer up to layer $l$. Each token has two dimensions for divided space-time attention, i.e., $z(p,t)$, where $p$ is a spatial dimension and $t$ is the depth dimension in the volumetric patch. Each TimeSFormer encoding block contains a time attention layer and a space attention layer. During the time attention block, each patch token only attends to patches at the same spatial locations, while during space attention, each patch only attends to the patches from the same frame. $T[p,j,q]$ represents the attention of $z(p,j)$ to $z(p,q)$ from the previous layer during time attention layer, where $T$ are the time attention weights. $S[i,j,p]$ represents the space attention of $z(i,j)$ to $z(p,j)$ from the time attention layer, where $S$ are space attention weights. Combining the space and time attention, each volumetric patch token attends to all patches at every spatial location within the volume through a unique path. The combined space-time attention $W$:
\begin{equation}
    W[i,j,p,q] = S[i,j,p] * T[p,j,q]
\end{equation}

\section*{\textbf{Results}} 
\subsection*{Morphology preserving alignment}
Elastic registration reduced the registration error by 50\% compared to rigid registration, from 40.1 $\mu m$ to 20.8 $\mu m$ median distance. The proposed non-rigid registration improvement achieved a similar performance (20.1 $\mu m$) and reduced the risk of glandular shape deterioration, which was examined by manual evaluation.

\subsection*{GGG diagnosis}
\begin{figure*}[ht!]
\includegraphics[width=6.5in]{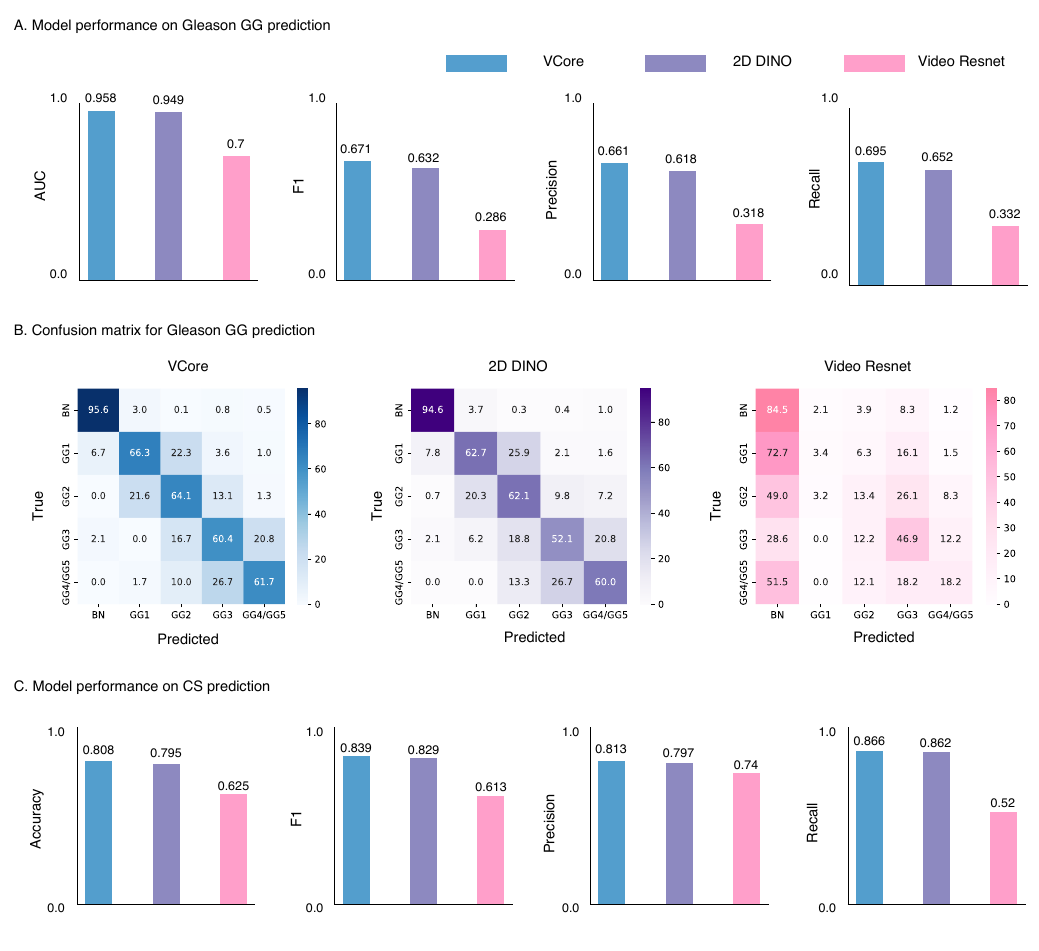}
\caption{\textbf{VCore Gleason GG classification results.}
(\textbf{a} AUC, F1, Precision and Recall for 5 class classification problem comparing volumetric, 2D and natural-image based features. ) 
(\textbf{b} Confusion matrix comparing actual and predicted labels for 5 class classification comparing volumetric, 2D, and natural-image-based features. ) 
(\textbf{c} AUC, F1, Precision and Recall for binary model performance (GGG $\ge$ 2 vs. GGG $<$ 2) comparing volumetric, 2D and natural-image based features. ) 
}
\label{fig3}
\end{figure*}

The results of GGG prediction are shown in FIG. \ref{fig3}A. Using volumetric features in ABMIL significantly outperformed (p $<$ 0.01) other feature types and achieved the highest AUC of 0.958, F1 of 0.671, Precision of 0.661, and Recall of 0.695. VCore resulted in a 1\% increase in AUC, a 6.5\% increase in the recall, a 6.1\% increase in precision, and a 5.8\% increase in the F1 score compared to the DINO framework, pretrained on 2D patches. A model trained on the video reset feature utilized in the previous 3D pathology work only obtained 0.7 AUC, 0.286 F1 score, 0.318 Precision, and 0.332 Recall, illustrating the importance of pathology-specific encoders.
FIG. \ref{fig3}B illustrates the confusion matrix for the VCore model in GGG prediction. VCore demonstrates superior performance across all classes (as depicted along the main diagonal) compared to the 2D DINO and Video Resnet models. VCore achieves higher accuracy in detecting GGG1 cases (66.3\% vs. 62.7\%) and a lower chance of classifying clinically insignificant (GGG1) as clinically significant (GGG2) (22.3\% vs. 25.9\%), which may influence treatment planning. The confusion matrix corresponding to Video Resnet indicates the model overfitting to the largest class (Benign).

Additionally, we binarized GGG labels and predictions as clinically significant (CS) (GGG $\ge$ 2) vs. not (GGG $<$ 2). We then trained the VCore framework on this dataset and found that VCore outperforms both 2D DINO and 2.5D Resnet baselines in accuracy (0.808 vs. 0.795 vs. 0.625), F1 score (0.839 vs. 0.829 vs. 0.613), precision (0.813, 0.797, 0.74) and recall (0.866, 0.862, 0.52) as shown in FIG. \ref{fig3}C.

\begin{figure*}[ht!]
\includegraphics[width=6.5in]{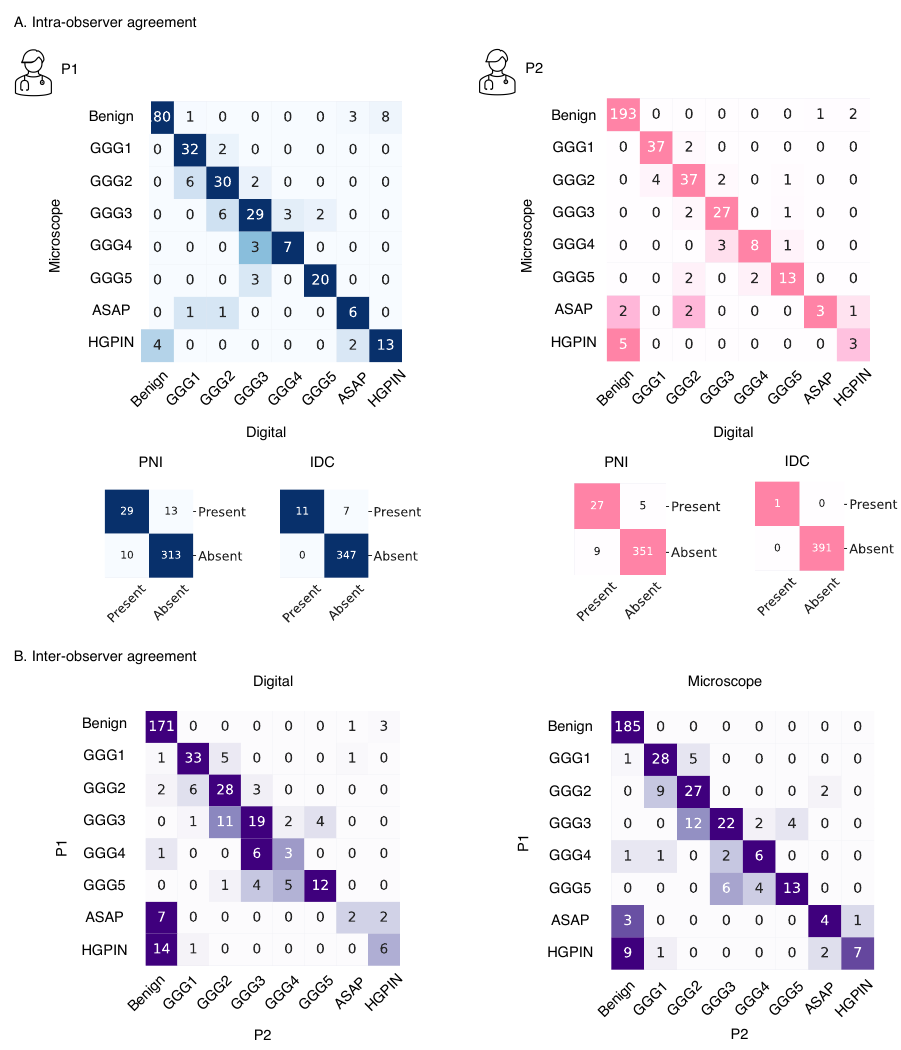}
\caption{\textbf{
Results of the clinical validation of the volumetric core using a digital pathology slide viewer with advanced functionality.}
(\textbf{a} Confusion matrix illustrating intra-observer variability for two pathologists (microscope assessment vs. digital assessment using the volumetric core)
(\textbf{b} Confusion matrix illustrating inter-observer variability between two pathologists (microscope assessment and digital assessment using the volumetric core)). ASAP - atypical small acinar proliferation,  HGPIN - high-grade prostatic intraepithelial neoplasia, PNI - perineural invasion, IDC - intraductal carcinoma.
}
\label{fig4}
\end{figure*}

\begin{figure*}[ht!]
\includegraphics[width=6.5in]{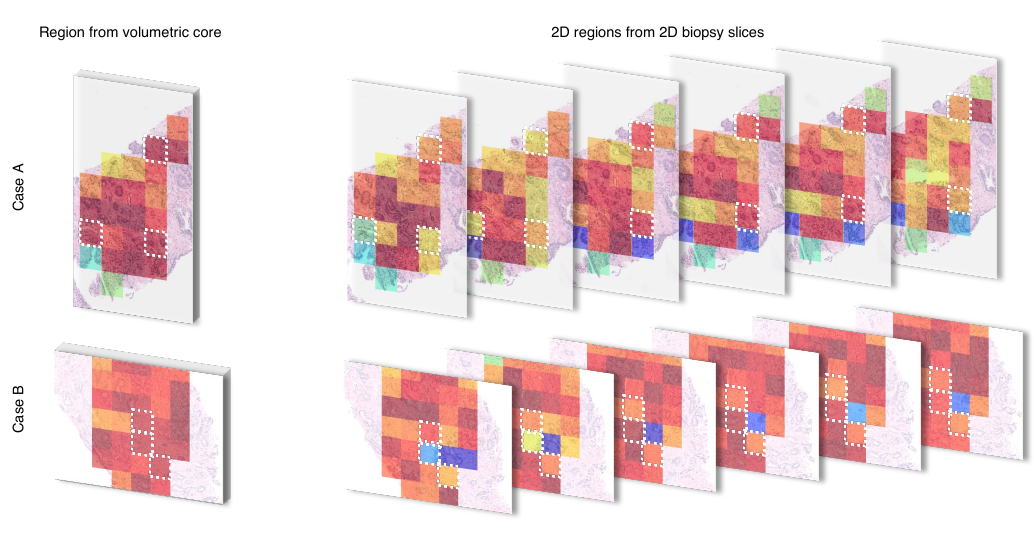}
\caption{\textbf{Visualization of the discriminative regions from WSIs overlaid with attention maps from ABMIL module from two clinically significant cores (Core A and Core B) where VCore resulted in a true positive (TP) prediction and 2D DINO resulted in a false negative (FN) prediction. White box indicates patches highlighted by pathologists.}
}
\label{fig5}
\end{figure*}

\begin{figure*}[ht!]
\includegraphics[width=6.5in]{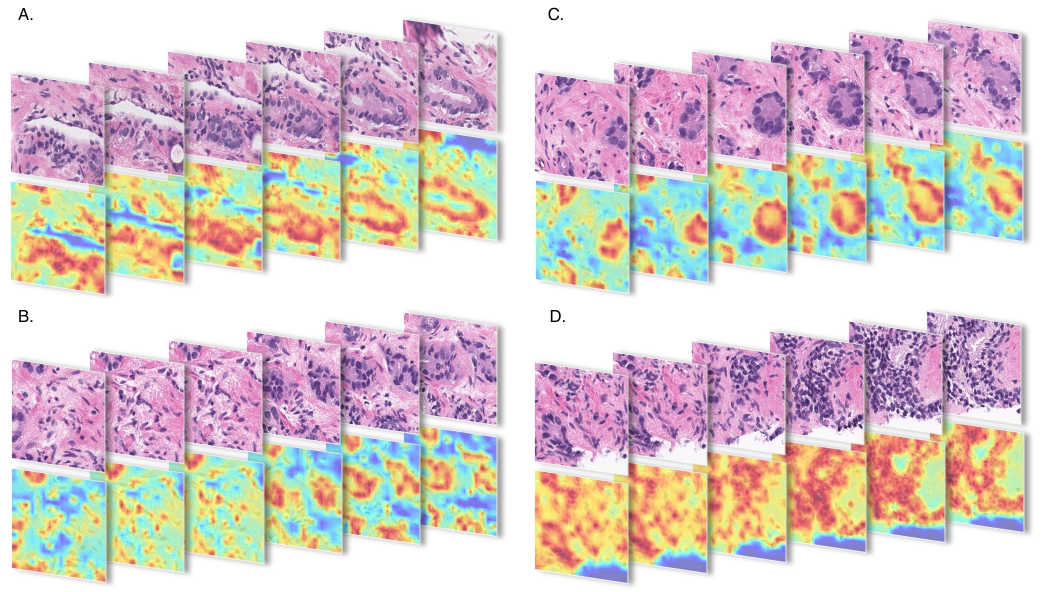}
\caption{\textbf{Visualization of the discriminative patches from WSIs overlaid with attention maps from the TimeSFormer model determined to be tumor patches containing volumetric information important to detecting clinical significance.}
}
\label{fig6}
\end{figure*}

The results show that a volumetric morphology-aware computational framework, which offers a natural approach to analyzing intrinsically 3D biological structures, has the potential to enhance diagnostic accuracy of automatic solutions. We support our argument by visualizing two discriminative regions from WSIs overlaid with attention maps from two clinically significant cores (see FIG. ~\ref{fig5}) correctly classified by VCore and misclassified by other models. Highlighted patches examined by pathologists were determined to be tumor patches containing volumetric information important to detecting clinical significance. 
With only a single 2D view of the tumor within these patches, there is a risk of either missing it or grading it inaccurately. In FIG. ~\ref{fig6} we overlay the highlighted patches with attention maps from TimeSFormer backbone, highlighting important glandular variations within the volume essential for correct grading.

\subsection*{Clinical validation}
We evaluate the intra-and inter-pathologists' agreement using quadratic Cohen Kappa $\kappa$, comparing diagnostic differences between traditional microscope-based and digital assessments utilizing the volumetric core. The inter-pathologists' agreement estimation resulted in $\kappa=0.7$ for microscope-based examination and $\kappa=0.73$ for digital examination 
utilizing a volumetric core. The improved $\kappa$ value for digital examination indicates better agreement among pathologists, reflecting increased consistency and reliability in the diagnostic process.   The intra-pathologist agreement estimation resulted in $\kappa=0.81$ for P1 and 
$\kappa=0.87$ for P2 comparing microscope-based to digital signout, indicating substantial agreement. 

A confusion matrix was used to compare both sign-out results on a core-by-core basis (see FIG. \ref{fig4}A).
While most cases identified as benign through microscopic examination remained classified as benign after digital examination using a volumetric core, a few cases were reclassified as HGPIN or ASAP by both pathologists. Only a few cases initially deemed clinically insignificant by microscopic examination were reclassified as clinically significant by both pathologists (5.8\% by P1 and 5.1\% by P2). Conversely, a higher proportion of cases initially identified as clinically significant through microscopic examination were subsequently classified as clinically insignificant after digital sign-out (15.7\% by P1 and 9.3\% by P2).

A confusion matrix was used to compare the diagnoses of P1 and P2 on a core-by-core basis (see FIG. \ref{fig4}B). 14.7\% of cores graded as GGG1 by P1 were graded as GGG2 (CS) by P2 during the microscope-based examination and 12.5\% during the digital examination utilizing the volumetric core. 23.6\% of cores graded as GGG2 by P1 were graded as GGG1 or benign by P2 during the microscope-based examination and 20.5\% during the digital examination utilizing the volumetric core.

\section*{\textbf{Discussion}} 
We present a computational platform to derive a novel data source from routinely scanned H\&E slides. Additionally, we introduce a DL-based algorithm that uses this new data type and outperforms existing solutions in determining the GGG grade of prostate biopsy. Given a cohort of routine slices, stained and scanned consecutive section tissue sections, our VCore approach can perform automatic morphology-preserving alignment and combine a feature encoder with an attention-based aggregation network to render patient-level predictions. The proposed framework could be applicable to other serial section data types, e.g. breast or renal biopsies.
The size of the collected cohort significantly exceeds that of 3D pathology datasets collected using non-destructive 3D imaging technologies, enabling the use of self-supervised pre-training for volumetric tissue patch feature encoders.

We first demonstrated the successful application of VCore for the GGG classification of prostate cancer using a large cohort generated by a morphology-preserving alignment framework. A significant performance drop was seen when using feature extractors pretrained on natural images, highlighting the importance of pathology-specific pretraining frameworks. Additionally, VCore showed higher performance compared to traditional 2D approaches. This result suggests that the proposed pretraining framework, which operates on volumetric patches, provides additional value over conventional 2D-based pretraining approaches.

In terms of clinical translation, a significant advantage of this technology is that it uses data generated from routinely scanned biopsy slides, making it non-destructive and enabling extensive data generation. Future studies could explore the developed AI tool's utility for scrolling through volumetric core biopsies from various cancers and how they can guide pathologists' decision-making.


An impactful benefit of using digital examination with the volumetric core is that it resulted in a smaller diagnosis discrepancy between P1 and P2 compared to microscope-based examination. The reduced discrepancy between pathologists ensures more consistent and reliable diagnoses, which is crucial for improved patient management and treatment planning.
As shown in the results, a lower percentage of disagreement occurred between pathologists within clinically significant and insignificant cases when using digital examination with the volumetric core. These results show promise for improving diagnostic accuracy, leading to more accurate differentiation between cases that require treatment and those that can be managed conservatively (e.g., using active surveillence)~\citep{tosoian2016active}.

\subsection*{\textbf{Limitations}} 
While involving two pathologists in this study was sufficient to demonstrate the potential of digital sign-out using a volumetric core, a larger number of clinicians would be needed to explore its potential fully. The larger study would enable the assessment of variability in diagnostic accuracy and interobserver consistency across a diverse range of clinical expertise. 

Expanding the study to a multicenter setting is crucial for evaluating the digital evaluation process using a volumetric core. The multicenter study could introduce variability in patient population and diagnostic workflows, which would help assess VCore's efficacy across different settings. 

Prostate cancer biopsies were chosen as a primary subject for this study because they typically involve the collection of the largest number of cores, often ranging from 10 to 20 or more, which makes pathological assessment a time-consuming and labor-intensive task. This extensive sampling is essential to accurately diagnose prostate cancer due to its heterogeneity. 
Nevertheless, the study should be fully extensible to other tissue types (e.g., breast or renal biopsy), where the number of cores is smaller but accurate evaluation of tissue structures within a biopsy volume is essential for diagnosis.
Additionally, future studies could extend the computational pipeline to other clinical tasks, such as survival prediction.

Our future work includes incorporating novel feature-aggregation methods built to improve attention-based MIL.
Improved aggregation methods that try to learn dependencies between features are crucial for patient-level prediction tasks.

Future directions could also include the discovery of volumetric morphological biomarkers that cannot be detected by 2D methods, enhancing various clinical and research applications. Although VCore framework can localize regions associated with specific clinical outcomes through the attention-based module of the classification network, 3D segmentation may be necessary to extract more explainable morphological features to improve diagnostic and prognostic performance. 

\section*{\textbf{Conclusion}}
Volumetric core presents a novel data source to improve AI-based tools to diagnose prostate biopsy cores automatically. Spatial morphology information in the x, y, and z directions is learned by a self-supervised pretraining framework based on
DINO with TimeSformer backbone, which leads to improved cancer grading accuracy.  
Additionally, a digital slide viewer with an advanced option to scroll through consecutive and co-registered tissue sections of a volumetric core helps pathologists track structures across multiple slices of a biopsy core, which leads to higher inter-rater agreement. Overall, the volumetric core represents a critical advancement in digital pathology, offering substantial benefits for clinical practice.

\section*{\textbf{Acknowledgments}}
The authors would like to acknowledge support from the NIH/NCI
R01CA279666. 
\section*{Competing interests}
The authors declare no competing interests.

\section*{Data availability}
The data that supports the plots within this paper are available from the corresponding author upon reasonable request.

%
\bibliographystyle{apsrev4-1}
\bibliography{reference}

\
\end{document}